\documentstyle[aps,epsf]{revtex}
\begin{document}
\title{Parameterization of the Equation of State and the expansion history of the very recent universe}
\author{
W. F. Kao\thanks{%
wfgore@cc.nctu.edu.tw} \\
Institute of Physics, Chiao Tung University, Hsinchu, Taiwan}
%\date{June 6, 2003}
\maketitle

\begin{abstract}
Motivated by a nice result shown by E. Linder, a detailed
discussion on the choice of the expansion parameters of the
Maclaurin series for the equation of states of a perfect fluid is
presented in this paper. We show that their nice recent result is
in fact a linear approximation to the full Maclaurin series as a
power series of the parameter $y=z/(1+z)$. The power series for
the energy density function, the Hubble parameter and related
physical quantities of interest are also shown in this paper. The
method presented here will have significant application in the
precision distance-redshift observations aiming to map out the
recent expansion history of the universe. In addition, a complete
analysis of all known advantageous parameterizations for the
equation of states to high redshift is also presented.
\end{abstract}


%\pacs{{\bf PACS\/}: }

%\preprint{hep-th/0306037}
\section{INTRODUCTION}
Recently there are great advances in our abilities in cosmological
observations for the quest of exploring the expansion history of
the universe. It carries cosmology well beyond determining the
present dimensionless density of matter $\Omega_m$ and the present
deceleration parameter $q_ 0$ of Sandage \cite{[1]}. One seeks to
reconstruct the entire function $a(t)$ representing the expansion
history of the universe. Earlier on, cosmologists sought only a
local measurement of the first two derivatives of the scale factor
a, evaluated at a single time $t_ 0$. One tries instead to map out
the function determining the global dynamics of the universe in
the near future. While many qualitative elements of cosmology
follow merely from the form of the metric, (see Weinberg
\cite{[2]}), deeper understanding of our universe requires
knowledge of the qualitative dynamics of the scale factor $a(t)$.
This echoes the transition of energy between components signifying
the epoch of radiation domination to that of matter domination.
This is also a key element in the growth of density perturbations
into large structure. Yet until recently the literature only
considered the Hubble constant and the deceleration parameter
measured today. There are now a great numbers of cosmological
observational tests that will be able to probe the function a(t)
more completely throughout all ages of the universe (see Sandage
\cite{[3]}, Linder \cite{[4]}, \cite{[5]}, Tegmark \cite{[6]}).

What one required is a probe capable of both precise and accurate
enough observations. Indeed, a number of promising methods are
being developed including the magnitude-redshift relation of Type
Ia supernovae. The goal of mapping out the recent expansion
history of the universe is well motivated. The thermal history of
the universe, extending back through structure formation,
matter-radiation decoupling, radiation thermalization, primordial
nucleosynthesis, etc. is very important in the study of cosmology
and particle physics, high energy physics, neutrino physics,
gravitational physics, nuclear physics, and so on (see, e.g., Kolb
and Turner \cite{[7]}).

Recent expansion history of the universe is similarly a very
promising research focus with the discovery of the current
acceleration of the expansion of the universe. This includes the
study of the role of high energy field theories in the form of
possible quintessence, scalar-tensor gravitation, higher dimension
theories, brane worlds, etc in the very recent universe. The
accelerated expansion is also important to the possible fate of
the universe \cite{[8]}, \cite{[9]}, \cite{[10]}. It is then
important to consider the use of supernova observations to obtain
the magnitude-redshift law back to $z \sim 1.7$ and how to relate
this to the scale factor-time behavior $a(t)$ with the proposed
Supernova/Acceleration Probe mission \cite{[11]}.

Therefore, the study of modelling different equation of state
(EOS) derived from different theories plays an important role in
the study of the very recent expansion history of our universe.
Interestingly, there is a nice paper showing a new model that is
capable of extracting important physics with the help of the
linear parameterization in powers of an appropriate parameter
$z/(1+z)$ expanding the function $\omega$ of the equation of
state\cite{[12]}, \cite{[13]}, \cite{[14]}, \cite{[15]}.  We will
show in section II that one should treat the proposed new model as
the leading term in the Maclaurin series associated with the EOS
(cf. \cite{[14]} and Fig. 1 of \cite{[15]}). In section III, we
will present the first four leading terms in computing the energy
density function $\rho$, the Hubble parameter $H$, and the
conformal time $\eta$ with the help of the Maclaurin series of the
EOS.

 \section{Taylor expansion of the Equation of State}
The equation of state is given by the relation $\omega = p/\rho$.
In addition, the field equations of the universe can be shown to
be:
\begin{eqnarray} \label{eos}
d \ln \rho (z)  &=& 3 (\omega +1) d \ln (1+z), \\
H^2 &=& {8 \pi G \over 3} \rho, \label{H}
\end{eqnarray}
for a flat universe. Here $1+z$ is the redshift defined by $a/a_0
= 1/(1+z)$. Various linear models \cite{[12]}, \cite{[13]} have
been suggested for the EOS in the literatures. We will try to
point out in this paper the underlying physics behind these linear
models and go further to study a more practical leading-orders
expansion.

Taylor expansion is known to be one of the best ways to extract
leading contributions from a generic theory with the help of a
power series expansion of some suitable field variables. The
Taylor series is normally convergent due to the structure of the
expansion coefficients. The power series will, however, converge
quickly if the range of variable is properly chosen. Hence it is
important to choose an appropriate expansion variable for the
purpose of our study.

For example, one may expand the EOS, assuming to be a smooth
function for all $z$, as a power series of the variable $z$ around
the point $z=0$. This will lead to the power series expansion:
\begin{equation}\label{z}
\omega(z)= \sum_n {\omega^{(n)}(z=0) \over n!} z^n .
\end{equation}
Here the summation of $n$ runs from $0$ to $\infty$ and
$f^{(n)}(z)={d^n f(z) \over dz^n}$. We will not specify clearly
the range of summation unless it is different from the one we
adopted. This convention will be applied to all the summation
ranges through out this paper for convenience.

Alternatively, one can
also expand the EOS as a power series of the variable $w=1/(1+z)$ around
the point $w(z=0)=1$. The result is
\begin{equation} \label{wz}
\omega(w(z))= \sum_n {\omega^{(n)}(w=1) \over n!} (w-1)^n = \sum_n
{\omega^{(n)}(w=1) \over n!} (-)^n ({z \over 1+z})^n .
\end{equation}

Note that if we expand the EOS in power of the variable
$y=z/(1+z)$ around $y(z=0)=0$, we will end up with a similar power
series:
\begin{equation} \label{yz}
\omega(y(z))= \sum_n {\omega^{(n)}(y=0) \over n!} y^n .
\end{equation}

The Taylor series or the Maclaurin series are normally convergent
due to the structure of the expansion coefficients. Nonetheless,
one would prefer to choose a more appropriate expansion parameter
in order to make the series converge more rapidly. As a result,
leading order terms will be enough to extract the most important
physics from the underlying theory. Therefore the advantage of the
$y$ expansion is that the power series converges rapidly for all
range of the parameter $-1< y=z/(1+z) <1$, or equivalently,
$-1/2<z< \infty$ as compared to the range $|z|<1$ for the power
series expansion of $z$ shown in Eq. (\ref{z}).

One notes, however, that one should also expand all physical
quantities and the field equations to the same order of precision
we adopted for the EOS expansion. Higher order contributions will
not be reliable unless one can show that the higher order terms
does not affect the physics very much. For example, the liner
order is good enough for the expansion of the EOS modelling SUGRA
model \cite{[13]}. This is because that the linear term fits the
predicted EOS for SUGRA model to a very high precision. One
readily realizes, however, that it is not easy to track the series
expansion order by order due to the form of the Eq. (\ref{eos})
for the EOS for the $y$ expansion. This is because that after
performing the integration, one needs to pay attention to the
distorted integration result. Indeed, one needs to write $y=1-w$
in order to perform the integration involving $d \ln w$. Indeed,
one will need to recombine the result back to a power series of
$y$. The trouble is that the lower order terms could sometimes
hide as the higher order terms in $w$. It may not be easy to track
clearly the lower order $y$ expansion when we perform the
expansion, for example, due to the exponential factor in the Eq.
(\ref{eos}). Therefore, one finds that the most natural way to
expand the EOS is to expand it in terms of the variable $x= \ln
(1+z)$ around the point $x(z=0)=0$. Indeed, this power series can
be shown to be:
\begin{equation} \label{x}
\omega(x(z))= \sum_n {\omega^{(n)} \over n!} x^n.
\end{equation}
And this power series converge very rapidly in the range $-1< \ln
(1+z) <1$, or equivalently, $-0.63 \sim 1/e - 1 < z <e -1 \sim
1.72$. Here $e \sim 2.72 $ is the natural factor. Note that this
limit happens to agree with the proposed scope of the SNAP
mission. We will study these different power series expansion for
the EOS and its applications in the following sections.

\section{Power series of $z/(1+z)$}

We will show in details how to extract the leading terms in the
$y$-expansion with $y=({z \over 1+z})$ around the point
$y(z=0)=0$. One can write the expansion coefficient as $\omega_n =
{(1+\omega)^{(n)}(y=0) \over n!}$ such that the power series for
the expansion of the EOS becomes
\begin{equation} \label{y}
1+\omega(y(z))= \sum_n \omega_n y^n .
\end{equation}
Hence the Eq. (\ref{eos}) can be shown to be
\begin{eqnarray} \label{eos1}
d \ln \rho (y)  = 3 (1+\omega)  { d y \over 1-y}= d\,\, \left[ \sum_n \sum_k 3
\omega_n {y^{n+k+1} \over n+k+1} \right ] .
\end{eqnarray}
Note that we are now expanding with respect to the smooth function
$(1+\omega)$, instead of $\omega$, as a Maclaurin series for
convenience. Therefore, one can integrate above equation to obtain
\begin{equation} \label{eos2}
\rho (y) = \rho_0 \exp [\,\, 3 \sum_n \sum_k {\omega_n y^{n+k+1}
\over n+k+1}\,\,] \equiv \rho_0 X(y)= \rho_0 \sum_n X_n y^n.
\end{equation}
Note that one needs to expand the function $\rho$ as a power
series of $y$ too in order to extract the approximated solution
with appropriate order. The expansion coefficient $X_n$ is defined
as $X_n = X^{(n)}(y=0)/n!$. Here, the superscript in $X'$ denotes
differentiation with respect to the argument $y$ of the function
$X(y)$. One can show that $X'=XY$ with $Y=3 \sum_n \sum_k \omega_n
y^{n+k}$. In addition, one can show that
\begin{equation}\label{Yl0}
Y^{(l)}(y)=3 \,\, \sum_{n} \sum_{k}{ (n+k)! \over (n+k-l)!}
\omega_n y^{n+k-l}.
\end{equation}
Hence one has
\begin{equation}\label{Yl}
Y^{(l)}(y=0)=3(l!) \,\, \sum_{n=0}^{l} \omega_n .
\end{equation}

One can also show, for example, that
\begin{eqnarray}
X''&=& X(Y^2+Y') \\
X'''&=& X(Y^3+3YY'+Y'') \\
X^{(4)} &=& X(Y^4+6Y^2Y'+3(Y')^2+4 YY''+Y''' ).
\end{eqnarray}
This series does not appear to have a more compact close form for
the multiple differentiation with respect to $y$. One can,
however, put the equations as a more compact format:
\begin{equation}
X^{(l+1)}=X [Y+{d \over dy}]^l Y.
\end{equation}
It appears, however, that one needs to do it manually even it is
straightforward. We will only list the leading terms as this is
already suitable expansion for our purpose at this moment.

Hence one has
\begin{eqnarray}
X_0&=&1, \\
X_1&=& 3 \omega_0, \\
X_2&=&{1 \over 2} [9 \omega_0^2+3\omega_0+3
\omega_1], \\
X_3&=& {1 \over 6} [27 \omega_0^3+27 \omega_0 (\omega_0+\omega_1)
+ 6 (\omega_0 + \omega_1 + \omega_2) ], \\
X_4 &=& {1 \over 24} [81 \omega_0^4+162 \omega_0^2
(\omega_0+\omega_1) + 27 (\omega_0 + \omega_1)^2 +
72\omega_0(\omega_0 + \omega_1 + \omega_2)+ 18(\omega_0 + \omega_1
+ \omega_2 +\omega_3) ].
\end{eqnarray}
Therefore, one can expand the final expression for the energy
density $\rho$ accordingly. Indeed, the result is
\begin{eqnarray}
\rho&=&\rho_0 \{  1 +  3 \omega_0 y + {1 \over 2} [9
\omega_0^2+3\omega_0+3 \omega_1] y^2 +  {1 \over 6} [27
\omega_0^3+27 \omega_0 (\omega_0+\omega_1)
+ 6 (\omega_0 + \omega_1 + \omega_2) ] y^3  \nonumber \\
&+& {1 \over 24} [81 \omega_0^4+162 \omega_0^2 (\omega_0+\omega_1)
+ 27 (\omega_0 + \omega_1)^2 + 72\omega_0(\omega_0 + \omega_1 +
\omega_2)+ 18(\omega_0 + \omega_1 + \omega_2 +\omega_3) ]y^4  \}
\end{eqnarray}
to the order of $y^4$. Note that one keep the order of precision
to $y^4$ in computing the energy density $\rho$ even we are
expanding the EOS only to the order of $y^3$. This is due to the
special structure in the energy momentum conservation law
\ref{eos}. In addition, one can show that the Hubble parameter
$H=H_0X^{1/2}$ with $H_0=\sqrt{8\pi G \rho_0 /3}$. And the
expansion for $X^{1/2}$ can be obtained by replacing all
$\omega_n$ with $\omega_n/2$ in writing the expansion for $X$.
Therefore one has
\begin{eqnarray}
H&=&H_0 \{  1 +  {3 \over 2} \omega_0 y + {1 \over 8} [9
\omega_0^2+6(\omega_0+ \omega_1)] y^2 +  {1 \over 48} [27
\omega_0^3+54 \omega_0 (\omega_0+\omega_1)
+ 24 (\omega_0 + \omega_1 + \omega_2) ] y^3  \nonumber \\
&+& {1 \over 192} [162 \omega_0^4+162 \omega_0^2
(\omega_0+\omega_1) + 54 (\omega_0 + \omega_1)^2 +
144\omega_0(\omega_0 + \omega_1 + \omega_2)+ 72(\omega_0 +
\omega_1 + \omega_2 +\omega_3) ]y^4  \}.
\end{eqnarray}

Note also that one can also compute the conformal time according
to the expression:
\begin{equation}
H_0 \eta = \int_0^z dz'X^{-{1 \over 2}}=\int dy {X^{-{1 \over 2}}
\over (1-y)^2 }.
\end{equation}
Knowing that $1/(1-y)^2=\sum_n (n+1)y^n$, one can show that
\begin{equation}
H_0 \eta = \int dy X^{-{1 \over 2}} \sum_n (n+1)y^n .
\end{equation}
Therefore, one can easily compute the expansion of $\eta$ in a
straightforward manner.

\section{power series of $\ln (1+z) $}
We will show in details how to extract the leading terms in the
$x$-expansion with $x=\ln (1+z)$ around the point $x(z=0)=0$. One
can write the expansion coefficient as $\omega_n =
{(1+\omega)^{(n)}(x=0) \over n!}$ such that the power series for
the expansion of the EOS becomes
\begin{equation} \label{xx}
1+\omega(x(z))= \sum_n \omega_n x^n .
\end{equation}
Note that we use the same notation for $\omega_n$ in different
parameterizations for convenience. Hence the Eq. (\ref{eos}) can
be shown to be
\begin{eqnarray} \label{eos1x}
d \ln \rho (x)  = 3 (1+\omega)  d x= d\,\, \left[ \sum_n 3
\omega_n {x^{n+1} \over n+1} \right ] .
\end{eqnarray}
Note that we are now expanding the physical quantities with
respect to the function $(1+\omega)$ instead of $\omega$ for
convenience. Therefore, one can integrate above equation to obtain
\begin{equation} \label{eos2x}
\rho (x) = \rho_0 \exp [\,\, 3 \sum_n  {\omega_n x^{n+1} \over
n+1}\,\,] \equiv \rho_0 X(x)= \rho_0 \sum_n X_n x^n.
\end{equation}
Note that one needs to expand the function $\rho$ as a power
series of $x$ too in order to extract the approximated solution
with appropriate order. The expansion coefficient $X_n$ is defined
as $X_n = X^{(n)}(x=0)/n!$. One can show that $X'=XY$ with $Y=3
\sum_n  \omega_n x^n=3(1+\omega)$. Therefore, one has
\begin{equation}\label{xlx}
Y^{(l)}(y=0)=3(l!) \,\, \omega_l .
\end{equation}

One can also show, for example, that
\begin{eqnarray}
X''&=& X(Y^2+Y') \\
X'''&=& X(Y^3+3YY'+Y'') \\
X^{(4)} &=& X(Y^4+6Y^2Y'+3(Y')^2+4 YY''+Y''' ).
\end{eqnarray}
In addition, one can show that
\begin{equation}
X^{(l)}= \sum_n \sum_k 3 X_n \omega_k { (n+k)! \over (n+k-l+1)!}
x^{n+k-l+1}.
\end{equation}
Therefore, one has
\begin{equation}
X_0^{(l)}= \sum_{n+k=l-1>0} 3 X_n \omega_k (l-1)!.
\end{equation}
Hence one obtains the recurrence relation for the expansion
coefficients of $X_n$:
\begin{equation}
X_l = {3 \over l} \sum_{n=o}^{l-1} \sum_k  X_n \omega_k .
\end{equation}
As a result, one has, for example,
\begin{eqnarray}
X_0&=&1, \\
X_1&=& 3 \omega_0, \\
X_2&=&{1 \over 2} [9 \omega_0^2+3
\omega_1], \\
X_3&=& {1 \over 6} [27 \omega_0^3+27 \omega_0 \omega_1
+ 6 \omega_2 ], \\
X_4 &=& {1 \over 24} [81 \omega_0^4+162 \omega_0^2
\omega_1 + 27 \omega_1^2 +
72\omega_0 \omega_2+ 18\omega_3 ].
\end{eqnarray}
Therefore, one can expand the final expression for the energy
density $\rho$ accordingly. Indeed, one has
\begin{eqnarray}
\rho&=&\rho_0 \{  1 +  3 \omega_0 y + {1 \over 2} [9 \omega_0^2+3
\omega_1] y^2 +  {1 \over 6} [27 \omega_0^3+27 \omega_0 \omega_1
+ 6  \omega_2 ] y^3  \nonumber \\
&+& {1 \over 24} [81 \omega_0^4+162 \omega_0^2 \omega_1 + 27
\omega_1^2 + 72\omega_0\omega_2+ 18\omega_3 ]y^4 \}.
\end{eqnarray}
In addition, one can show that the Hubble parameter $H=H_0X^{1/2}$
with $H_0=\sqrt{8\pi G \rho_0 /3}$. And the expansion for
$X^{1/2}$ can be obtained by replacing all $\omega_n$ with
$\omega_n/2$ in writing the expansion for $X$. The result is
\begin{eqnarray}
H&=&H_0 \{  1 +  {3 \over 2}  \omega_0 y + {1 \over 8} [9
\omega_0^2+6 \omega_1] y^2 +  {1 \over 48} [27 \omega_0^3+54
\omega_0 \omega_1
+ 24  \omega_2 ] y^3  \nonumber \\
&+& {1 \over 384} [81 \omega_0^4+324 \omega_0^2 \omega_1 + 108
\omega_1^2 + 288\omega_0\omega_2+ 144\omega_3 ]y^4 \}.
\end{eqnarray}
Note also that one can also compute the conformal time according
to the expression:
\begin{equation}
H_0 \eta = \int_0^z dz'X^{-{1 \over 2}}=\int dy {X^{-{1 \over 2}}
\over (1-y)^2 }.
\end{equation}
Knowing that $1/(1-y)^2=\sum_n (n+1)y^n$, one can show that
\begin{equation}
H_0 \eta = \int dy X^{-{1 \over 2}} \sum_n (n+1)y^n .
\end{equation}
Therefore, one can easily compute the expansion of $\eta$ in a
straightforward manner.

\section{conclusion}
The proposed Supernova/Acceleration Probe (SNAP) will carry out
observations aiming to determine the components and equations of
state of the energy density, providing insights into the
cosmological model, the nature of the accelerating dark energy,
and potentially clues to fundamental high energy physics theories
and gravitation. As a result, we are motivated by a nice result
shown in Ref. \cite{[13]} to study the physics underlying the
model presented there.

A detailed discussion on the choices of the expansion parameters
of the Maclaurin series for the equation of states of a perfect
fluid is presented in this paper accordingly. For example, the
Maclaurin series of the EOS is expanded as power series of the
variables $y=z/(1+z)$ and $x=\ln (1+z)$ respectively. We also show
how to obtain the power series for the energy density function,
the Hubble parameter and related physical quantities of interest.
The method presented here will have significant application in the
precision distance-redshift observations end to map out the recent
expansion history of the universe, including the present
acceleration and the transition to matter dominated deceleration.
We also show that the nice recent result in Ref. \cite{[13]} is in
fact a linear approximation to the full Maclaurin series as a
power series of the parameter $y=z/(1+z)$.

Since we can power expand all smooth EOS into a convergent power
series, it is more practical for the future probe to measure local
derivatives of the EOS. Note that $\omega_0$ represents the value
of $1+\omega$ at time $t_0$ or $z=0$. In addition, $\omega_1$ is
the slope of the curve $\omega$ vs $y$ or $\omega$ vs $x$ at
$z=0$. Therefore, knowing the values of the expansion coefficients
will enable us to plot the entire EOS through out the range of
convergence. One may need to use different expansion series
depending on the convergent speed of the power series. For
example, it appears that the leading order term in the $y$
expansion is good enough to obtain a nice result for the SUGRA
prediction. This is because the leading term is close enough to
the theoretical prediction \cite{[13]}. Nonetheless, one expects
that a few leading terms in the Maclaurin series will be able to
offer close enough result for us in the future.

In addition, the result shown here is independent of the choice of
the time $t_0$. The local measurement of the expansion
coefficients can be extended to compare the expansion coefficient
at any time. One probably should trust less on the results other
than the local predictions.

\section*{Acknowledgments}

%%%%%%%%%%%%%%%%%%%%%%%%%%%%%%%%%%%%%%%%%%%%%%%%%%%%%%%%%%%%%%%%%%%%%%
This work is supported in part by the National Science Council
under the grant numbers NSC91-2112-M009-034.
%\begin{references}

\end{document}